\documentclass[12pt,prd,showkeys,showpacs]{revtex4}
\usepackage[dvipdfm]{graphicx,color} 
\usepackage{amsmath}
\numberwithin{equation}{section}

\begin{document}

\begin{titlepage}

\begin{flushright}
OIQP-12-01
\end{flushright}

\vspace{\baselineskip}

\begin{center}
\large \bf Pysical Account of Weyl Anomaly from Dirac Sea\\
\end{center}

\vspace{1.5\baselineskip}

\begin{center}
Y.\ Habara${}^{a}$, H.\ B.\ Nielsen${}^{b}$ and M.\ Ninomiya${}^{a}$
\end{center}

\vspace{\baselineskip}

\begin{center}
\it
${}^{a}$Okayama Institute for Quantum Physics,\\
Kyo-yama 1-9-1  Kitaku, Okayama 700-0015, Japan
\end{center}

\vspace{0.5\baselineskip}

\begin{center}
\it
${}^{b}$Niels Bohr Institute, University of Copenhagen,\\
17 Belgdamsvej, DK 2100 Denmark
\end{center}

\vspace{2\baselineskip}
\begin{abstract}

We derive the Weyl anomaly in two dimensional space-time by considering the Dirac sea regularized by some negatively counted formally bosonic extra species. 
In fact we calculate the trace of the energy-momentum tensor of the Dirac sea in a background gravitational field. It has to be regularized, since otherwise the Dirac sea is bottomless and thus causes divergence. The new regularization method consists in adding various massive species some of which are to be counted negative in the Dirac sea. The mass terms in the Lagrangian of the regularization fields have a dependence on the background gravitational field. 
\end{abstract}


\maketitle
\end{titlepage}

\section{INTRODUCTION}\label{sec:intro}

In 1973 Capper and Duff~\cite{1}~\cite{2}~\cite{3}~\cite{4} discovered that the correlation function of 
the energy-momentum tensor for a massless particle theory 
\begin{equation}
\Pi _{\mu \nu \rho \sigma }(p)
=\int d^{u}xe^{ipx}\textless T_{\mu \nu }(x) T_{\rho \sigma }(o)\textgreater 
| _{g_{\mu \nu } =\delta _{\mu \nu }}
\label{1.1}
\end{equation}
although it obeyed as classically expected the conservation of 4-momentum Ward-identities
\begin{equation}
p^{\mu }\Pi _{\mu \nu \rho \sigma }(p)=0
\end{equation}
does not obey the expected tracelessness of the energy-momentum tensor, since indeed the finite part are not ``traceless''  i.e.\
\begin{equation}
\Pi ^{\mu }_{\> \> \mu \rho \sigma }(\mbox{finite part})\neq 0.
\end{equation}

Classically of course one expects that for a theory of massless particles the energy-momentum tensor should be traceless $T^{\mu }_{\> \> \mu }=0$ and so obtaining $T^{\mu }_{\> \> \mu }\neq 0$ in any correlation function signals an \underline{anomaly}.

In the present work we hope to throw some light onto the question of how such an anomaly comes about physically. We have chosen as our technology to extract the $T^{\mu }_{\> \> \mu }$ which is the Hamiltonian even if there is no interaction between the massless particles but they just are in some gravitational non-trivial background that can act on $T_{\rho \sigma }(x)$ there appears this effect of the trace of the $T_{\mu \nu }$ not being zero. This effect is referred to as the Weyl-anomaly.

Really the reason for this $T^{\mu }_{\> \> \mu }\neq 0$ being an ``anomaly'' is that a theory with only massless particles formally classically obey symmetry under scaling and that $T^{\mu }_{\> \> \mu }=0$ can be extracted from the requirement of scaling invariance (Dilatation symmetry = scaling symmetry.)

The generator for dilatation symmetry is 
\begin{equation}
D=\int X^{\mu }T_{\mu 0}d^{3}\vec{x}
\end{equation}
or the current for this $D$ is 
\begin{equation}
j_{D \mu }(x)=x^{\rho}T_{\rho \mu }(x)
\end{equation}

In fact we shall use as this background --- which could act on the $T_{\mu \nu}$ ---
\begin{equation}
\eta _{\mu \nu }\longrightarrow g_{\mu \nu}=\eta ' _{\mu \nu}
=e^{\Omega }\eta _{\mu \nu }
\end{equation}
where $\Omega $ has been used as an ansatz for the Weyl transformation function.

In the present article we shall only perform our calculation in 1+1 dimensions --- but with a Euclideanized technique so that we are really using 2+0 dimentions ---.

When we now consider the vacuum to vacuum transition in the by some Weyl transformation modified.

In the following section 2 fermion and Dirac Sea~\cite{5} we shall introduce some notation for fermions in a background gravitational field obtained from that space-time by a Weyl transformation and we shall describe the Dirac sea. In \underline{section 3} we then prepare for how to extract the vacuum expectation value of the trace $T^{\mu }_{\> \> \mu }$ of the energy momentum tensor by varying the vacuum to vacuum $S$-matrix element in a background metric $e^{2\Omega }\eta _{\mu \nu}$. In \underline{section 4} we describe our cut-off procedure by cutting-off the (negative) high energy part of the Dirac sea by compensating its contribution to say, energy and momentum by means of a ghost-like particles, which is really a massive boson following a fermion like equation of motion. Using this cut-off procedure we compute in \underline{section 5} the correlation function for the Weyl transformation background field $\Omega (x)$ which according to section 3 is required to achieve for instance the Weyl anomaly expression for $T^{\mu }_{\> \> \mu}$.

In section 6 we then conclude and outlook resume.

\section{NOTATIONS AND DIRAC SEA}\label{sec:notation}

The model we shall consider is a two component --- thus Dirac --- fermion in a gravitational field background which is though assumed to be conformally flat.

It is in reality given by a Weyl transformation having acted on a flat space-time. I.e.\ the metric is of the form 
\begin{equation}
g_{\mu \nu }=e^{2\Omega }\eta _{\mu \nu }.
\end{equation}
Here $\eta _{\mu \nu }$ is the flat metric. We then consider a two component complex fermion 
\begin{equation}
\Psi =\begin{pmatrix} \overline{\psi} \\ \psi \end{pmatrix}
\end{equation}
on the Euclidean 2 dimensional space with coordinates \ $\vec{x}=(x^{0}, x^{1})=(t, x)$, so that \\ $-\infty < t < \infty $ \ and \ $0 \le  x^{1} < 2\pi $. We in fact require periodicity with period $2\pi$ and consider the cylindrical space-time $S^{1} \times R$.

In the present paper we adopt notations such that Roman indices $i,j,\cdots$ take the component of the flat space while Greek ones $\mu , \nu \cdots $ take those of the curved space. The diffeomorphism invariant action reads 
\begin{eqnarray}
S&=&\frac{1}{2\pi }\int d^{2}\vec{x} \sqrt{g} \Psi ^{+}(\vec{x})\gamma ^{0}\gamma ^{i}e^{\mu}_{i}\times \bigtriangledown _{\mu}\Psi (\vec{x}) \nonumber\\
&=&\frac{1}{4\pi }\int d^{2}\vec{x} \sqrt{g} \{ \Psi ^{+}(\vec{x})\gamma ^{0}\gamma ^{i}\times e^{\mu}_{i}(\vec{x}) \bigtriangledown _{\mu}\Psi (\vec{x}) \nonumber\\
&-&\bigtriangledown _{\mu }\Psi ^{+}(\vec{x})\gamma ^{0}\gamma ^{i}e^{\mu}_{i}(\vec{x}) \Psi (\vec{x})\} 
\label{2.3}
\end{eqnarray}

By making use of the diffeomorphism the metric tensor $g_{\mu \nu }$ can be made into conformal flat form 
\begin{equation}
g_{\mu \nu }=e^{2\Omega (\vec{x})}\eta _{\mu \nu }
\end{equation}

Hereafter we assume that as $t \to  \pm \infty $ space-time becomes flat, i.e.\ $ \lim_{t \to {\pm \infty} } g_{\mu \nu} = \eta _{\mu \nu }$ and 
\begin{equation}
\lim_{t \to {\pm \infty} } \Omega (\vec{x}) = 0.
\label{2.5}
\end{equation}
Since we deal with conformal transformation we introduce the complex coordinate
\begin{eqnarray}
z&=&e^{x+ix^1}, \nonumber\\
\overline{z}&=&e^{x^0-ix^1}.
\end{eqnarray}
That purpose we need to introduce zweibeins $e^{\mu }_{i}(x)$ --- or their inverses $f^{i}_{\mu }(x)$ ---. We may choose to specify these zweibeins to be diagonal as a ``gauge choice'' so that 
\begin{equation}
e^{0}_{1}=e^{1}_{0}
\end{equation}
and the requirement
\begin{equation}
\eta _{ij}e^{i}_{\mu }e^{j}_{\nu }=g_{\mu \nu }=e^{2\Omega }\eta _{\mu \nu}
\label{2.8}
\end{equation}
would then lead to the choice
\begin{equation}
e^{1}_{1}=e^{0}_{0}=e^{\Omega }
\end{equation}
and 
\begin{equation}
f^{1}_{1}=f^{0}_{0}=e^{-\Omega }
\end{equation}
for the inverse $f^{\mu }_{i}$.

Then we can write the action
\begin{eqnarray}
S&=&\frac{1}{2\pi }\int d^{2}\vec{x} \sqrt{g} \Psi ^{+}(\vec{x})\gamma ^{0}\gamma ^{i}e^{\mu}_{i}\times \bigtriangledown _{\mu}\Psi (\vec{x}) \nonumber\\
&=&\frac{1}{4\pi }\int d^{2}\vec{x} \sqrt{g} \{ \Psi ^{+}(\vec{x})\gamma ^{0}\gamma ^{i}\times e^{\mu}_{i}(\vec{x}) \bigtriangledown _{\mu}\Psi (\vec{x}) \nonumber\\
&-&\bigtriangledown _{\mu }\Psi ^{+}(\vec{x})\gamma ^{0}\gamma ^{i}e^{\mu}_{i}(\vec{x}) \Psi (\vec{x})\}. 
\end{eqnarray}

We may rewrite this action into the form
\begin{eqnarray}
S&=&\frac{1}{4\pi }\int d^{2}\vec{x} e^{2\Omega }\cdot 
\Bigl\{(\overline{\psi }^{+}, \psi )(\vec{x})e^{-\Omega }
\begin{pmatrix} \bigtriangledown _{2}+i\bigtriangledown _{1} & 0 \\ 0 &  \bigtriangledown _{2}-i\bigtriangledown _{1} \end{pmatrix}
\begin{pmatrix} \overline{\psi } \vec{x} \\  \psi \vec{x} \end{pmatrix} \nonumber\\
&-&(\bigtriangledown _{2}+i\bigtriangledown _{1})\overline{\psi} ^{+}\vec{x}e^{-\Omega }\overline{\psi} (\vec{x}) \nonumber\\
&-&(\bigtriangledown _{2}-i\bigtriangledown _{1}) \psi e^{-\Omega }\psi 
(\vec{x}) \Bigr\} \nonumber\\
&=&\frac{1}{4\pi }\int d^{2}\vec{x}\Bigl(\overline{\psi }^{+} e^{\Omega }(\bigtriangledown _{2}+i\bigtriangledown _{1})\overline{\psi } \nonumber\\
&+& \psi ^{+}e^{\Omega }(\bigtriangledown _{2}-i\bigtriangledown _{1})\psi \nonumber\\
&-&\bigl((\bigtriangledown _{2}+i\bigtriangledown _{1})\overline{\psi }^{+}\bigr) e^{+\Omega }\overline{\psi } \nonumber\\
&+& (\bigtriangledown _{2}-i\bigtriangledown _{1})\psi ^{+} e^{\Omega }\psi \Bigr)
\label{2.12}
\end{eqnarray}

Here we have used the Weyl representation for $\gamma ^{\mu }$-matrices
\begin{eqnarray}
\gamma ^{1}=\begin{pmatrix} 0 & i \\ -i & 0 \end{pmatrix} \nonumber\\
\gamma ^{2}=\begin{pmatrix} 0 & -1 \\ -1 & 0 \end{pmatrix}
\end{eqnarray}
and the above expressions of zweibeins and metric in terms of $\Omega (\vec{x})$:
\begin{eqnarray}
f^{\mu }_{i}&=&\delta ^{\mu }_{i}e^{-\Omega } \nonumber\\
e^{i}_{\mu }&=&\delta ^{i}_{\mu }\cdot e^{\Omega } \nonumber\\
\mbox{and}\ \ g_{\mu \nu}&=&e^{2\Omega }\eta _{\mu \nu}.
\end{eqnarray}

We may rewrite the action (\ref{2.12}) into the form
\begin{eqnarray}
S&=&\frac{1}{4\pi }\int d^{2}\vec{x} \Bigl[ \bigl\{e^{\frac{1}{2}\Omega}\psi ^{+} \bigr\} 
(\bigtriangledown _{2}+i\bigtriangledown _{1})(e^{\frac{1}{2}\Omega}\psi)\nonumber\\
&-&(\bigtriangledown _{2}+i\bigtriangledown _{1})
(e^{\frac{1}{2} \Omega}\psi^{+})
\times (e^{\frac{1}{2} \Omega}\psi )+ \bigl\{e^{\frac{1}{2}\Omega} \overline{\psi} ^{+} \bigr\}
(\bigtriangledown _{2}-i\bigtriangledown _{1})\bigl\{e^{\frac{1}{2}\Omega }\overline{\psi}\bigr\} \nonumber\\
&-&(\bigtriangledown _{2}-i\bigtriangledown _{1}) (e^{\frac{1}{2} \Omega}\overline{\psi}^{+} )e^{\frac{1}{2} \Omega}\overline{\psi}\Bigr]
\label{2.15}
\end{eqnarray}

Here the prefactors of $\psi $ and $\overline{\psi }$ are
\begin{eqnarray}
e^{\frac{1}{2} \Omega} \nonumber\\
\mbox{and} \ \ e^{\frac{1}{2} \Omega}
\end{eqnarray}
respectively.

In this form the action is immediately seen to possess invariance under further Weyl transformation given by
\begin{eqnarray}
\Omega (\vec{x}) \to \Omega'  (\vec{x}) = \Omega (\vec{x}) + E(\vec{x})\nonumber\\
\psi (\vec{x}) \to \psi' (\vec{x}) \to e^{-\frac{1}{2} E(\vec{x})}\psi (\vec{x}) \nonumber\\
\overline{\psi } (\vec{x}) \to \overline{\psi }(\vec{x})' = e^{-\frac{1}{2} E(\vec{x})} \overline{\psi }(\vec{x})
\end{eqnarray}

Here $E(\vec{x})$ is the function describing the Weyl transformation.

Noticing that in the equation (\ref{2.15}) the derivatives from Leibnitz rule acting on $\Omega $ drop out (cancel) we see that the equation of motion would be like if $\Omega $ were constant and in any case we find the equations of motion
\begin{eqnarray}
(\bigtriangledown _{2}+i\bigtriangledown _{1})(e^{\frac{1}{2}\Omega}\psi)=0\nonumber\\
(\bigtriangledown _{2}-i\bigtriangledown _{1})(e^{\frac{1}{2}\Omega}\psi ^{+})=0\nonumber\\
(\bigtriangledown _{2}-i\bigtriangledown _{1})(e^{\frac{1}{2}\Omega}\overline{\psi })=0\nonumber\\
(\bigtriangledown _{2}-i\bigtriangledown _{1})(e^{\frac{1}{2}\Omega}\overline{\psi} ^{+})=0
\end{eqnarray}
if you vary independently $\psi, \ \psi ^{+}, \ \overline{\psi } \  \mbox{and} \ \overline{\psi} ^{+}$. In any case you easily get by defining the ``tilded'' fields
\begin{eqnarray}
\widetilde{\psi } (\vec{x})&=&e^{\frac{1}{2}\Omega (\vec{x}) }\psi (\vec{x}) \nonumber\\
\widetilde{\psi }^{+} (\vec{x})&=&e^{\frac{1}{2}\Omega (\vec{x}) }\psi^{+}  (\vec{x}) \nonumber\\
\widetilde{\overline{\psi }}(\vec{x})&=&e^{\frac{1}{2}\Omega (\vec{x}) }\overline{\psi } (\vec{x}) 
\nonumber\\
\widetilde{\overline{\psi }}^{+}(\vec{x})&=&e^{\frac{1}{2}\Omega (\vec{x}) }\overline{\psi }^{+} (\vec{x}) 
\label{2.19}
\end{eqnarray}
the (seemingly) $\Omega $-independent action
\begin{eqnarray}
S&=&\frac{1}{4\pi }\int d^{2}\vec{x} \Bigl[ \widetilde{\psi }^{+}(\bigtriangledown _{2}+i\bigtriangledown _{1})\widetilde{\psi }-\bigl( (\bigtriangledown _{2}+i\bigtriangledown _{1})\widetilde{\psi }^{+}\bigr)\cdot \widetilde{\psi }\nonumber\\
&+&\widetilde{\overline{\psi }}^{+} (\bigtriangledown _{2}-i\bigtriangledown _{1})\widetilde{\overline{\psi }}-\bigl( (\bigtriangledown _{2}-i\bigtriangledown _{1})\widetilde{\overline{\psi }}^{+}\bigr)\cdot \overline{\psi }\Bigr].
\label{2.20}
\end{eqnarray}
The a priori covariant derivatives $\bigtriangledown_{\mu }$ are when acting on the effective scalar --- the fermion $\psi , \ \overline{\psi }$ etc.\ and $\Omega $ --- just the usual derivative (operators) with respect to 
\begin{equation}
x^{\mu }, \ \bigtriangledown_{\mu }\sim \partial_{\mu}, \\ \bigtriangledown _{2}+i\bigtriangledown _{1}=\partial_{2}+i\partial_{1} \ \mbox{etc}.
\end{equation}
If we choose as new variables
\begin{eqnarray}
\ln{z}&=&x^{2}+ix^{1} \nonumber\\
and \ \ \ln{\overline{z}}&=&x^{2}-ix^{1}
\label{2.22}
\end{eqnarray}
and then
\begin{eqnarray}
\bigtriangledown _{2}+i\bigtriangledown _{1}&=&\frac{\partial}{\partial \ln{z}} \nonumber\\
\bigtriangledown _{2}-i\bigtriangledown _{1}&=&\frac{\partial}{\partial \overline{\ln{z.}}}
\end{eqnarray}
The equations of motion for $\widetilde{\overline{\psi }}$ and $\widetilde{\psi }$ becomes that they \emph{only} depend on respectively $\ln{z}=x^{2}+ix^{1}$ and $\ln{\overline{z}}=x^{2}-ix^{1}$.

We simply find right and left mover fields respectively $\widetilde{\psi }$ and $\widetilde{\overline{\psi }}$ and with the compactification of the space coordinate $x^{1}$ to have period
\begin{equation}
x^{1}\simeq x^{1}+2\pi
\end{equation}
We get the quantization of momentum $p^{1}$ and thus energy to
\begin{equation}
p^{1}=n \ \ \ (n:\mbox{integer}) \nonumber
\end{equation}

In fact
\begin{equation}
E=\pm p^{1}
\end{equation}

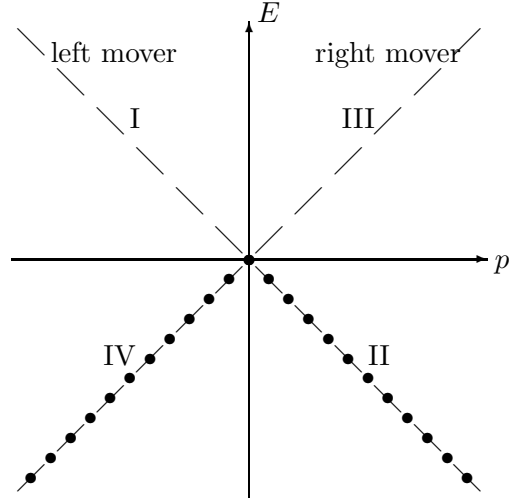
\begin{figure}[htb]
\begin{center}
\begin{picture}(200,190)
	\put(10,100){\vector(1,0){180}}
	\put(100,10){\vector(0,1){180}}
	\put(193,97){$p$}
	\put(103,190){$E$}
	\multiput(12.5,12.5)(15,15){12}{\line(1,1){10}}
	\multiput(12.5,187.5)(15,-15){12}{\line(1,-1){10}}
	\multiput(100,100)(-7.5,-7.5){12}{\circle*{4}}
	\multiput(100,100)(7.5,-7.5){12}{\circle*{4}}
	\put(135,150){III}
	\put(45,60){IV}
	\put(55,150){I}
	\put(145,60){II}
	\put(25,175){left mover}
	\put(125,175){right mover}
\end{picture}
\end{center}
\caption[Dirac Sea]{The particle distribution of the Dirac Sea} \label{fig:1} \end{figure}

Filling the Dirac Sea as here formulated in $\widetilde{\psi }$ , $\widetilde{\overline{\psi }}$-notation thus looks simply like filling states in Fig.\ref{fig:1} into the negative energy parts of the two line-dispersion laws.

According to say equation (\ref{2.12}) the canonically conjugate to say equation  $\overline{\psi }$ is
\begin{equation}
\frac{\partial L}{\partial (\partial_{2}\overline{\psi })}
=\frac{1}{4\pi }e^{\Omega }\overline{\psi }^{+}
\end{equation}
where
\begin{eqnarray}
L&=&\frac{1}{4\pi }\bigl( \overline{\psi }^{+}e^{\Omega }(\bigtriangledown _{2}+i\bigtriangledown _{1})\overline{\psi }\bigr)+{\psi }^{+}e^{\Omega }(\bigtriangledown _{2}-i\bigtriangledown _{1}){\psi }\nonumber\\
&-&\bigl( (\bigtriangledown _{2}+i\bigtriangledown _{1})\overline{\psi }^{+}\bigr)e^{\Omega }\overline{\psi }+(\bigtriangledown _{2}-i\bigtriangledown _{1}){\psi }^{+}e^{\Omega }{\psi }.
\end{eqnarray}

>From there we then obtain the anti-commutation relations for the second quantized fields
\begin{eqnarray}
\bigl\{ \psi (t, x^{1}), \psi (t, {x'^{1}})\bigr\} &=&e^{\Omega(t, -x^{1}) }\delta (x^{1}-{x'^{1}})\nonumber\\
\bigl\{ \overline{\psi} (t, x^{1}), \overline{\psi} (t, {x'}^{1})\bigr\}&=&e^{\Omega(t, x^{1}) }\delta (x^{1}-{x'}^{1})
\end{eqnarray}

These anti-commutation rules match with the $\Omega$-independent rules for the tilted fields $\widetilde{\Psi },\ \widetilde{\overline{\Psi }} $ where they are defined from equation (\ref{2.19})
\begin{eqnarray}
\widetilde{\Psi }(\vec{x})&=&e^{\frac{1}{2}\Omega (\vec{x})}{\Psi }(\vec{x})\nonumber\\
\widetilde{\overline{\Psi }}(\vec{x})&=&e^{\frac{1}{2}\Omega (\vec{x})}\overline{\Psi }(\vec{x})\nonumber\\
\widetilde{\Psi }^{+}(\vec{x})&=&e^{\frac{1}{2}\Omega (\vec{x})}{\Psi }^{+}(\vec{x})\nonumber\\
\widetilde{\overline{\Psi }}^{+}(\vec{x})&=&e^{\frac{1}{2}\Omega (\vec{x})}\overline{\Psi }^{+}(\vec{x}),
\end{eqnarray}
namely
\begin{eqnarray}
\bigl\{ \widetilde{\Psi }(t, x^{1}), \ \widetilde{\Psi }^{+}(t, {x'}^{1})\bigr\}&=&\delta (x^{1}-{x^{1}}')\nonumber\\
\bigl\{ \widetilde{\overline{\Psi }}(t, x^{1}), \ \widetilde{\overline{\Psi }}^{+}(t, {x'}^{1})\bigr\}&=&\delta (x^{1}-{x'}^{1}).
\end{eqnarray}

In the completely usual way we may expand these second quantized fields on annihilation and creation operators $b_{n}$ and $b^{+}_{n}$ for momentum $p^{1}=n$,
\begin{eqnarray}
e^{-\frac{1}{2}\Omega (t,x)}\Psi (t,x^{1}) 
=\widetilde{\Psi }(t,x^{1})&=&\sum _{n=-\infty}^{\infty}\widetilde{b}_{n}e^{(-x^{2}-ix^{1})\cdot n} \nonumber\\
&=&\sum _{n=-\infty}^{\infty}\frac{\widetilde{b}_{n}}{z^{n}}
\end{eqnarray}
and
\begin{eqnarray}
e^{-\frac{1}{2}\Omega (t,x)}\overline{\Psi} (t,x^{1})
=\widetilde{\overline{\Psi }}(t,x^{1})&=&\sum _{n=-\infty}^{\infty}\widetilde{\overline{b}}_{n}e^{(-x^{2}-ix^{1})\cdot n} \nonumber\\
&=&\sum _{n=-\infty}^{\infty}\frac{\widetilde{\overline{b}}_{n}}{z^{n}}
\end{eqnarray}
Here the $\widetilde{b}_{n}$ and $\widetilde{\overline{b}}_{n}$ have the usual anti-communication relations
\begin{eqnarray}
\bigl\{ \widetilde{b}_{n},\widetilde{b}^{+}_{m} \bigr\}=\delta _{nm}\nonumber\\
\bigl\{ \widetilde{\overline{b}}_{n},\widetilde{\overline{b}}^{+}_{m} \bigr\}=\delta _{nm}
\end{eqnarray}
They anti-commute if we ask $b$ with $b$ rather than with $b^{+}$ or $\{b,\overline{b}^{(+)}\}=0$.

A priori we should put a bracket with $\Omega$ i.e.\ $[\Omega]$ onto all these creation and annihilation operators $\widetilde{b}^{+}_{n}$, $\widetilde{\overline{b}}^{+}_{n}$, $\widetilde{b}_{n}$ and $\widetilde{\overline{b}}_{n}$ so as to write rather $\widetilde{b}^{[\Omega]+}_{n}$, $\widetilde{\overline{b}}^{+[\Omega]}_{n}$, $\widetilde{b}^{[\Omega]}_{n}$ and $\widetilde{\overline{b}}^{[\Omega]}_{n}$.  However, since they have the same properties and could be identified if we insisted it is not really needed. In this creation notation the second quantized Hamiltonian comes to look
\begin{eqnarray}
H&=&\frac{1}{4\pi }\sum _{n=-\infty}^{\infty}\bigl( n+\frac{1}{2}\bigr) \bigl( \widetilde{b}^{+[\Omega]}_{n} \widetilde{b}^{[\Omega]}_{n}-\widetilde{b}^{[\Omega]}_{n} \widetilde{b}^{[\Omega]+}_{n} + \widetilde{\overline{b}}^{+[\Omega]}_{n}\widetilde{\overline{b}}^{[\Omega]}_{n}-\widetilde{\overline{b}}^{[\Omega]}_{n}\widetilde{\overline{b}}^{[\Omega]+}_{n}\bigr).
\label{2.34}
\end{eqnarray}

Corresponding to this Hamiltonian we can then construct the Dirac Sea by filling the single particle states with negative energies. Using that for the ``right moving'' $\Psi (t,x^{1})$ or $\widetilde{\Psi} (t,x^{1})$ 
\begin{equation}
E=p^{1}
\end{equation}
while for the ``left moving'' $\overline{\Psi} (t,x^{1})$ or $\widetilde{\overline{\Psi}}(t,x^{1})$ we have
\begin{equation}
E=-p^{1}
\end{equation}
we construct the Dirac Sea vacuum as 
\begin{equation}
|\mbox{sea}>=\Pi _{n\ge 0}\widetilde{\overline{b}}^{[\Omega]+}_{n}|\overline{0}>\otimes \Pi _{m< 0}\widetilde{b}^{[\Omega]+}_{m}|0>1
\end{equation}
where $|\overline{0}>$ and $|0>$ represent the ``fundamental'' vacua in the bar and no bar sectors in which there is not even the Dirac Sea(s). (see Fig.\ref{fig:1})

\section{SIGNIFICANCE OF THE DIRAC SEA}\label{sec:significance}

It is the philosophy of the present article to think of e.g.\ vacuum expectation value of $T_{\mu \nu }(x)$ (in the vacuum with the Dirac sea) as being due to this Dirac sea. To get meaningful results for a Dirac sea it is however needed to regularize it in some way or another so as to obtain finite though cut-off dependent e.g.\ energy density rather than just divergence. 

The technique which we also describe in a slightly simplified form in the following section consists in inventing a series of massive particle species some of which are ``ghosts'' in the sense of being counted negatively when it comes to the constructions of, say, $T_{\mu \nu }$ from their Dirac seas
~\cite{6}~\cite{7}~\cite{8}. ``Negatively counted'' species may be a better name since ``ghosts'' is used for something similar but not exactly the same. That is to say that in our regularization procedure we introduce two series of extra species with different masses. Then the idea is to let some of these extra species count negatively --- in the sense that their contributions to energy momentum etc.\ (say particle number charge) are counted negatively--- while others are counted just as usual fermions.

The basic idea now is to arrange the masses for these extra species so as to cancel out the contributions from the numerically large single particle energies so that the combined system of species together with the original fermion gets cut off. The typical mass of the extra species come to function as the cut-off scale $\wedge $.

In order to get the main (quadratic) divergence field we need to have including the original fermion just equally many species ``counted negatively'' as counted positively.

In order to get the logarithmic divergence cancel it is needed to arrange that the coefficient to the term of the form $\frac{1}{n}$ in the large $n$ expansion of the energy $E_{n}$ of the momentum $n\sim p$ level for the various introduced extra species plus the original fermion cancel out. Since the energy $E_{n}$ of a particle with mass $m$ and momentum $n=p$ is large $n$ expanded as 
\begin{equation}
E_{n}=\pm \sqrt{m^{2}+n^{2}}\approx \pm \bigl(n+ \frac{m^{2}}{2n}t\ldots \bigr)
\end{equation}
the condition to be imposed to cut-off the ``logarithmic divergence'' --- meaning the coefficient in the total counting (the ``negative counted'' counted with an extra minus sign) to $\frac{1}{n}$ --- is
\begin{equation}
\sum _{\mathrm{species}\atop{\mathrm{``extra + original''}}}\pm m^{2}=0
\label{c1}
\end{equation}

Here of course the $\pm$ is --- for the ``negative'' and + for the original fermion and extra species being counted positively. In the same notation the main cut-off being indeed a cut-off condition is 
\begin{equation}
\Sigma _{\mathrm{species}\atop{\mathrm{``extra \, plus \, original''}}}\pm1=0.
\label{c2}
\end{equation}

With the conditions (\ref{c1} and \ref{c2}) the combined system will indeed function as a cut-off. 
Since the proposed  cut-off is just based on massive particles (counted through negatively some of them), it will be Lorentz invariant and translational invariant and particle number conserving. However, it will \underline{no more} have \underline{conformal invariance}, nor \underline{Weyl invariance}, nor chiral invariance!
With such a cut-off we should thus be able to preserve Lorentz invariance and translational invariance, but risk anomalies in Weyl and scale invariance.

That is to say that for a vacuum, which does not \underline{spontaneously} break the mentioned symmetries, we should find $<T_{\mu \nu}>{{\propto}\atop {\sim }} g_{\mu \nu} $. 

While in the uncut-off theory it looks so conformally invariant that the variation of $\Omega$ in the metric $e^{2\Omega }\eta_{\mu \nu }$ is not felt the fermion field, with our cut-off procedure such an influence can come in. Indeed the mass terms in the Lagrangian for our extra species become $\Omega $-dependent. Indeed we have in the Weyl-transformation modified flat space-time metric  $g_{\mu \nu}=e^{2\Omega }\eta_{\mu \nu }$ the massive Dirac equation Lagrangian
\begin{equation}
\sqrt{g}L_{D}=\sqrt{g}\overline{\Psi }(x)(r^{\mu }e^{\mu }_{a}\partial _{\mu }-m)\Psi (x)
\end{equation}
so that even after going to the tilded notation $\widetilde{\Psi} (x)=e^{\frac{1}{2}\Omega }\Psi (x)$ we have the mass term
\begin{equation}
\sqrt{g}L=\ldots +\widetilde{\Psi }^{+}\gamma ^{0}m \widetilde{\Psi } e^{\Omega }.
\label{3.5}
\end{equation}

By expanding the exponential $e^{\Omega }$ this term gives rise to ``Yukawa''-like couplings of the ``extra species'' to $\Omega $.

If we do not care for the logarithmic divergence but only go for calculating the Weyl or conformal anomaly meaning $T^{\mu}_{\ \mu}$ we may not care to fulfill (\ref{c1}) and can if we like do with only one extra species, and that one should then be negatively counted. We shall do so in the following section 4.

\section{WEYL ANOMALY FROM DIRAC SEA}\label{sec:weyl anomaly}

\subsection{On how to extract $T^{\mu }_{\> \> \mu }$}\label{subsec:how to}

Since the Weyl or equivalently the conformal is known to mean that the trace of the energy momentum tensor~\cite{9} $T^{\mu }_{\> \> \mu }$ turns out to be nonzero, in fact $-\frac{1}{48\pi }R$ where $R$ is the Ricci scalar curvature, we need a procedure for extracting this energy momentum tensor $T^{\mu \nu}$.
It is well known [Birrel-Davies] that (interpreting $T_{\mu \nu}$ as renormalized are (see the footnote 2 lines above)) the expression in terms of the fields of a theory including a metric $g_{\mu \nu }$ for the energy momentum tensor $T_{{\mu }\nu}$ is obtainable from the action $S$ by functional differentiation with respect to the metric
\begin{equation}
T_{\mu \nu}=\frac{\delta S_{\scriptstyle(\mathrm{matter})}}{\sqrt{g}\delta g^{\mu \nu}}
\nonumber
\end{equation}
or
\begin{equation}
T_{\mu \nu}=-\frac{\delta S_{\scriptstyle(\mathrm{matter})}}{\sqrt{g}\delta g_{\mu \nu}}
\label{4.1}
\end{equation}
If we rather than the formal expression in terms of the fields such as $\Psi $ would like the expectation value in (say) the vacuum situation with some background gravitational field --- we think of our $g_{\mu \nu}=e^{2\Omega }\eta _{\mu \nu}$ above --- we might extract this expectation value for $T_{\mu \nu}(\vec{x})$ at a certain space-time point $\vec{x}$ by logarithmically functionally differentiating the vacuum to vacuum S-matrix / transition matrix element
\begin{equation}
<T_{\mu \nu }>=\frac{\delta <\mathrm{sea}|e^{-i\int _{-\infty}^{\infty}Hdt}|\mathrm{sea}>}{\delta g^{\mu \nu }}
\Big/<\mathrm{sea}|e^{-i\int _{-\infty}^{\infty}Hdt}|\mathrm{sea}>
\end{equation}

Here the vacuum in which we are interested should be --- of course --- the one \underline{with} the Dirac Sea filed. Here it should be understood that the second quantized Hamiltonian $H$ should contain the background metric $e^{2 \Omega }g_{\mu \nu }$ (unless it ``accidentally'' drops out). This means that \underline{a priori} the vacuum could develop away from being a vacuum --- getting e.g.\ pairs produced --- due to the effect of the background metric. However, as we have seen in the action (\ref{2.20}) formally our background field $\Omega $ and thus $g_{\mu \nu}=e^{2\Omega} \eta _{\mu \nu}$ does \underline{not} influence the fermions described in the $\widetilde{\psi },\widetilde{\overline{\psi }}$ notation at all. Thus unless the cut-off might change the situation, there is no effect of the considered background metric. Thus if this holds the Dirac Sea vacuum $|\mathrm{sea}>$ will remain undistributed by the background metric in the $\widetilde{\psi },\widetilde{\overline{\psi }}$ notation.

Let us remember though that it is only because we keep to the still \underline{conformally flat} metric $e^{2\Omega} \eta _{\mu \nu}$, that there is no effect of the background metric. Keeping to metric only being of the $e^{2\Omega} \eta _{\mu \nu}$ type we cannot extract the $T_{\mu \nu }$ proper, because we cannot vary the metric arbitrarily, but we \underline{can} extract the \underline{trace} $T^{\mu }_{\> \> \mu }=g^{\mu \nu }T_{\mu _\nu}=g_{\mu \nu }T^{\mu \nu }$, since indeed
\begin{eqnarray}
T^{\mu}_{\ \mu}&=& g_{\mu \nu}T^{\mu \nu} \nonumber\\
&=& g_{\mu \nu }\frac{\delta \ln <\mathrm{sea}|e^{-i\int _{-\infty}^{\infty}Hdt}|\mathrm{sea}>}{\delta g_{\mu \nu }} \nonumber\\
&=& \frac{1}{2}\frac{\delta \ln <\mathrm{sea}|e^{-i\int _{-\infty}^{\infty}Hdt}|\mathrm{sea}>}{\delta \Omega }.
\end{eqnarray}

Indeed we have of course with $g_{\mu \nu }=e^{2 \Omega }\eta _{\mu \nu}$ that 
\begin{eqnarray}
\frac{\delta \ln <\mathrm{sea}|e^{-i\int _{-\infty}^{\infty}Hdt}|\mathrm{sea}>}{\delta \Omega }
&=&\frac{\partial g_{\mu \nu } }{\partial \Omega}\Bigm|_{\vec{x}} \frac{\delta \ln <\mathrm{sea}|e^{-i\int _{-\infty}^{\infty}Hdt}|\mathrm{sea}>}{\delta g_{\mu \nu }}\nonumber\\
&=&2g_{\mu \nu }\frac{\delta \ln <\mathrm{sea}|e^{-i\int _{-\infty}^{\infty}Hdt}|\mathrm{sea}>}{\delta g_{\mu \nu }}\nonumber\\
&=&2g_{\mu \nu }T^{\mu \nu}=2T^{\mu}_{\ \mu}
\end{eqnarray}

We might thus extract the trace $T^{\mu }_{\> \> \mu }$ of the energy momentum tensor $T_{\mu \nu }$ alone from varying the background field by some (extra) Weyl transformation by say $\omega $ i.e.\
\begin{equation}
g_{\mu \nu }\to e^{2\omega }g_{\mu \nu }
\end{equation}
and looking for the variation of the $S$-matrix element from vacuum to vacuum.

Since we already wrote the formalism above for a by one Weyl transformation modified space, it may be most effective to just combine the further for $T^{\nu }_{\ \mu }$-extracting purposes introduced Weyl transformation $\omega $ with the already introduced one $\Omega $ to one combined
\begin{equation}
\Omega_{\mathrm{total}}=\Omega + \omega 
\label{4.6}
\end{equation}
Weyl transformation
\begin{equation}
\eta _{\mu \nu }\to e^{2 \Omega _{\rm{total}}} \eta _{\mu \nu }.
\end{equation}

\subsection{``Second quantized formalism''}\label{subsec:second}

Since in $\widetilde{\overline{\psi }},\widetilde{\psi }$-formulation we have effectively flat space --- only with an $S^{1}$-circle space, $R$x$S^{1}$ space-time --- in spite of a non-trivial $\Omega$ or say $\Omega _{\mathrm{total}}$ in (6.4), we can in reality in $\Omega _{\mathrm{total}}$-independent way expand the second quantized fermion fields $\widetilde{\overline{\psi }}$ and $\widetilde{\psi }$ and their hermitean conjugate annihilation and creation operators
\begin{equation}
\widetilde{\psi }(\vec{x})=\Sigma _{n \ \rm{integer}\atop{-\infty}}^{\infty}b_{n}e^{\rm{in}(x'-t)}
\label{4.8}
\end{equation}
and correspondingly the daggered second quantized fields would be expanded on a priori creation operators
\begin{eqnarray}
\widetilde{\psi }^{+}(\vec{x})&=&\Sigma _{n \ \rm{integer}\atop{-\infty}}^{\infty}\widetilde{b}^{+}_{n}e^{\rm{-in}(x'-t)}\\
\label{4.9}
\widetilde{\overline{\psi }}(\vec{x})&=&\Sigma _{n \ \rm{integer}\atop{-\infty}}^{\infty}\overline{\widetilde{b}}^{+}_{n}e^{\rm{in}(x'-t)}
\label{4.10}
\end{eqnarray}

Formulated in the complete space-time description using eq.\ (\ref{2.22}) inserted into the formulas (\ref{4.8}-{4.10}) we obtain the expansion for $\psi $'s.
\begin{eqnarray}
\widetilde{\psi }(z)&=&\Sigma _{n=-\infty}^{\infty}\widetilde{b}_{n} \frac{1}{z^{n+\frac{1}{2}}}\nonumber\\
\widetilde{\overline{\psi }}(\overline{z})&=&\Sigma _{n=-\infty}^{\infty}\widetilde{\overline{b}}_{n} \frac{1}{\overline{z}^{n+\frac{1}{2}}}
\end{eqnarray}

\subsection{The Dirac Sea}\label{subsec:dirac sea}

If we work in the Euclidean space-time, we have for the flat case --- or if we ignore as we can $\Omega _{\mathrm{total}}$ because it does not couple ---.

A priori we have a different world each time we change the background field $\Omega _{\mathrm{total}}$ and thus in principle we should have the creation and annihilation operators $\widetilde{b}^{+}_{n}$, $\widetilde{\overline{b}}^{+}_{n}$, $\widetilde{b}_{n}$ and $\widetilde{\overline{b}}_{n}$ depend on $\Omega _{\mathrm{total}}=\Omega + \omega $ so that we should write for example $\widetilde{b}^{[\omega ]}_{n}$, $\widetilde{\overline{b}}^{[\omega ]}_{n}$, $\widetilde{b}^{+[\omega ]}_{n}$ and $\widetilde{\overline{b}}^{[\omega ]}_{n}$.

However, since $\Omega_{\mathrm{total}}$ does \underline{not} appear in the equations of motion of the $\widetilde{\psi }$ and $\widetilde{\overline{\psi }}$-fields we may suggestively ignore such $\omega $-dependence and identify them such as
\begin{equation}
\widetilde{b}^{[\omega ]}_{n}=\widetilde{b}^{[0]}_{n}
\end{equation}
the energy operator $-\frac{d}{dt}$ while the momentum operator is $-i\frac{d}{dx^{'}}$

Thus the dispersion relation is depicted in Fig.\ref{fig:1}.

\section{OUR CUT-OFF PROCEDURE}\label{sec:cut-off}

We have seen above that formally the modification of the metric from $\eta _{\mu \nu}$ to $e^{2\Omega(x')}\eta _{\mu \nu}$ (see (\ref{2.3})) makes no change in the Hamiltonian (see (\ref{2.34})). Really this is seen (also) from the equations (\ref{2.34}), in which $H'$ has the same expression in $\widetilde{b}^{[\omega ]}_{n}$ and $\overline{\widetilde{b}}^{[\omega ]}_{n}$ as $H$ in $\widetilde{b}^{[0]}_{n}$ and $\overline{\widetilde{b}}^{[0]}_{n}$.

We can thus stress that in these (formal) expressions the modification of the metric has completely dropped out. So any dependence on the ``Weyl transformations'' on the flat space can only come in via the regularization.

In this section 5 we shall now propose a regularization of most importantly the Dirac sea.
Indeed, a regularization is performed by adding to the system yet a particle species in addition to the fermion described by $b^{[0]}_{n}$ or $b^{[\omega ]}_{n}$ (which are actually equal to each other). This added particle species should have the quantities to result from its Dirac sea be \underline{subtracted} rather than added as for our usual fermion. So we should declare that energy momentum and particle number from the Dirac sea for this added species should be counted with an extra minus sign.

Let us immediately give the idea that using our earlier works an ``Dirac sea for Bosons''~\cite{6}~\cite{7}~\cite{8} a boson with exactly the same action and equations of motion as the fermions we start from would have this property of subtraction its contribution from collected quantities such as energy momentum and particle number because we found out the Dirac sea for bosons should have one particle \underline{removed} (i.e.\ added -1 particle) from each negative energy single particle state. Thus we should imagine that our added species to cancel the contribution from the fermions could be a boson with exactly the same equation and spin etc.\ as the fermion. It would thus not obey the spin statistics theorem in general but rather remind of a ghost-particle species.

Now we do not want such a proposed ghost to cancel all the contribution from the fermion Dirac sea, but only the beyond the wanted cut-off part. We propose therefore the ``ghost particle'' --- the boson with fermionic equation of motion and ``spin'' --- to have a mass $M$ of the order of the wanted cut-off $\wedge $, i.e.\ $M \sim \wedge $. Using such a massive ``ghost'' has the advantage of letting the theory including the cut-off have the usual symmetries such as translational invariance and particle number conservation, but not Weyl invariance and not chiral symmetry.

So with this cut-off procedure we cannot get anomalies in momentum or particle number conservation, but we ``risk'' to get a Weyl anomaly as well as a chiral anomaly.

While the massless fermion is formally untouched by the modification $\Omega (x)$ of the metric $e^{2 \Omega(x)}\eta _{\mu \nu }$ the ghost-like boson to remove its high energy contribution will ``feel'' this modification via its mass $M$. Indeed the mass term should in principle be understood relative to the physical metric tensor $g_{\mu \nu}=e^{2\Omega}\eta_{\mu \nu }$.

We shall be allowed to use totally flat space-time as long as we do not cut-off and use the $\widetilde{\psi }$-fields. However, the mass term which brings the cut-off by being there for the compensating ghost-like species must physically be defined in the $\psi$ -notation.

For instance the mass term $M \overline{\psi}_{nc} \psi_{nc}$ for one of our  ``negatively counted'' fields in $\psi_{nc}$-notation would be written in the $\widetilde{\psi}_{nc}$-notation as analogous to (\ref{3.5})

\begin{eqnarray}
L_{\mathrm{mass}\ {nc}_{1}}&=&\wedge \overline{\psi }_{nc}(x)\psi_{nc}(x) \nonumber\\
&=&\wedge \overline{\widetilde{\psi}}_{nc}e^{-\frac{1}{2}\Omega(x)} e^{-\frac{1}{2}\Omega(x)} \widetilde{\psi}_{nc}(x)\nonumber\\
&=&\wedge \overline{\widetilde{\psi}}_{nc}(x)\widetilde{\psi}(x)e^{-\Omega(x)}
\end{eqnarray}
by using for $\psi_{nc}$ the analogous rewritting to a tilde field $\widetilde{\psi}_{nc}$ to (\ref{2.19}). Here we used the symbol $\wedge$ for the mass in the untilded notation to remind us that this mass is indeed a cut off. In the tilde-notation we then have an (effective) mass
\begin{equation}
M(x^{\mu})=\wedge e^{-\Omega (x)}
\end{equation}
which is now space-time dependent.

It is not hard to check that this $\Omega$-dependence of the effective mass $M(x^{\mu})$ is consistent with dimensionality considerations. In fact the distance element $ds$ is given by 
\begin{eqnarray}
ds^{2}&=&g_{\mu \nu}dx^{\mu}dx^{\nu} \nonumber\\
&=&e^{2\Omega}\eta_{\mu \nu }dx^{\mu}dx^{\nu} 
\end{eqnarray}
so that the physical distance $ds$ is $e^{\Omega}$ times the flat distance element $ds_{\rm{FLAT}}$ given by
\begin{equation}
ds^{2}_{\rm{FLAT}}=\eta_{\mu \nu}dx^{\mu}dx^{\nu} 
\end{equation}

So a mass $M$ given as say $\wedge$ in the $ds$ measuring would in the flat notation based on $ds_{\rm{FLAT}}$ look like being scaled opposite to the distance --- since $M$ has dimension of inverse distance --- 
\begin{equation}
M(x^{\mu})=\wedge e^{-\Omega}
\end{equation}

Of course such an $x^{\mu}$-dependent mass $M(x^{\mu})=\wedge e^{-\Omega(x^{\mu})}$ can be interpreted as an interaction of the Fermion (really the boson-ghost) in a Yukawa-type way with the background field $\Omega(x)$ by expanding the mass term 
\begin{equation}
M(x)\overline{\widetilde{\psi}}_{nc}\widetilde{\psi}_{nc}=\wedge \overline{\widetilde{\psi}}_{nc}\widetilde{\psi}_{nc}-\Omega(x)\wedge \overline{\widetilde{\psi}}_{nc}\widetilde{\psi}_{nc}+\ldots  
\end{equation}

The $\Omega$-dependence only come in via the higher order terms in this expansion firstly of course via
\begin{equation}
-\wedge \Omega \overline{\widetilde{\psi}}_{nc}\widetilde{\psi}_{nc}
\end{equation}
and there is no interaction with the original fermion field $\widetilde{\psi}$ itself, only with this ``compensating'' ghost-field $\widetilde{\psi}_{nc}$. ($\mbox{nc}$ stands for ``negatively counted.'')

To truly cancel even logarithmic divergenses we may need both positively and negatively counted massive extra particles such as $\psi_{nc}$. In fig 1.5 a suggestive picture of the dispersion relations for the originally occurring particles plus the extra species invented in order to regularize the energy and momentum from the Dirac sea. These dispersion relations are just ordinary relativistic dispersion laws, only the extra ``ghosts'' are massive while the original particle is massless.

\begin{figure}[htbp]
	\begin{center}
	\includegraphics[width=12cm,angle=270]{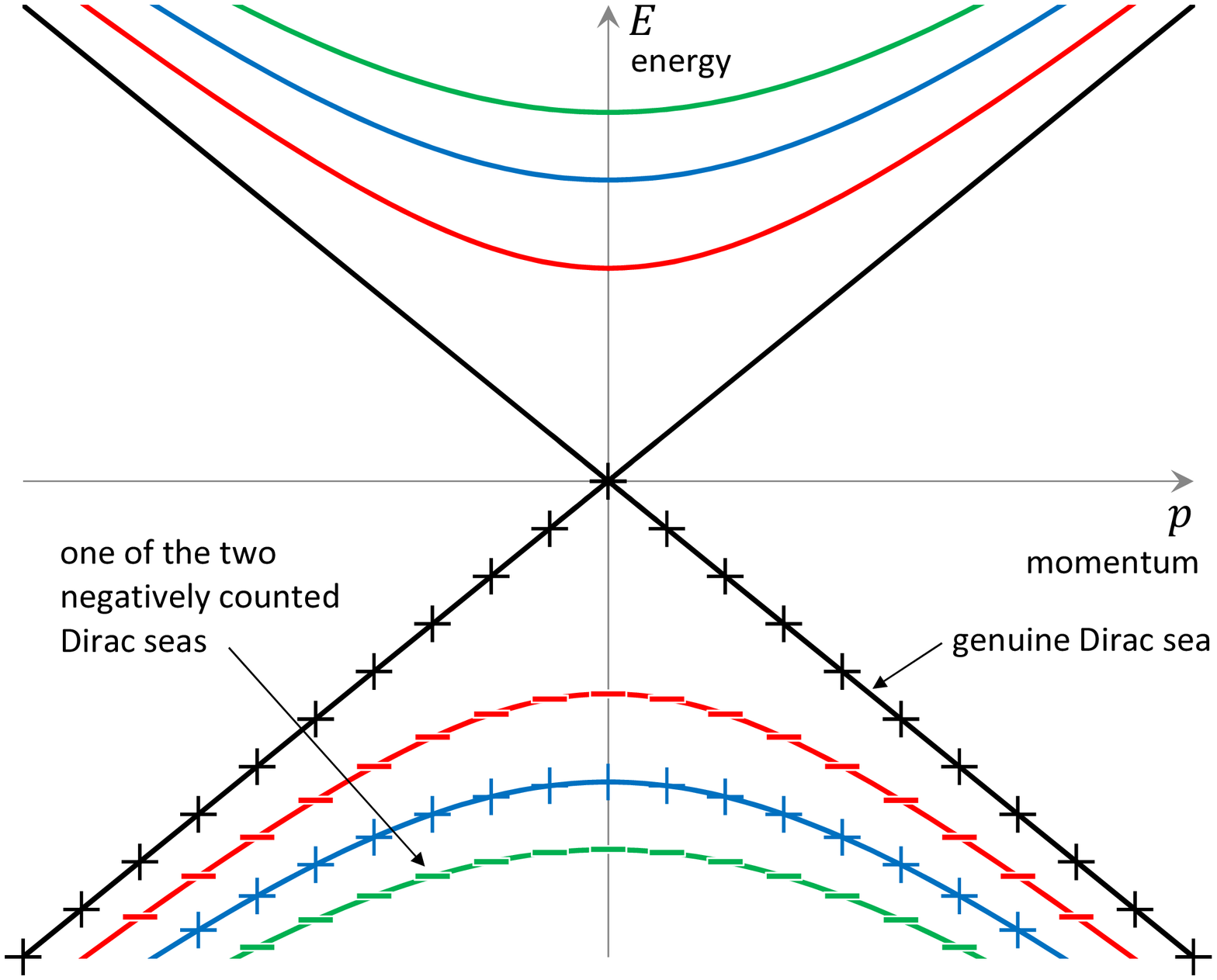}
	\end{center}
	\caption{This figure illustrates the dispersion relations for the original 
	fermion and the ``little series " of three for regularization purpose 
	introduced particle species, two of which are counted as there being a 
	negative number $-1$ of particles in the Dirac sea as indicated by the 
	small minussigns ``-" on the negative energy branches of the dispersion 
	relations. The four different species have their dispersion relations 
	denoted by respectively.}
	\label{fig:1.5}
\end{figure}

\section{CALCULATION OF TWO-POINT FUNCTION FOR $\omega$, THE WEYL TRANSFORMATION FUNCTION}\label{sec:culculation}

In order to evaluate the dependence of the functional derivative (\ref{4.1}) on a further Weyl transformation $\omega$ leading to the $\Omega_{\mathrm{total}}=\Omega+\omega$ (\ref{4.6}) we need to evaluate $ln <\mathrm{sea}|e^{-i\int _{-\infty}^{\infty}Hdt}|\mathrm{sea}>$ at least to second order in $\Omega_{\mathrm{total}}$. Now according to the discussion above --- see (\ref{2.20})--- the $\Omega_{\mathrm{total}}$ dependence come in only via the cut-off which means then the extra particle which in our simplified case is the negatively counted field. We shall now calculate the one-loop correction to the second order term in the background $\Omega_{\mathrm{total}}$ field. It is the idea here to do this by considering the one loop ``vacuum'' diagrams due to the $\Omega_{\mathrm{total}}$ interaction vertices.

\begin{figure}[htbp]
	\begin{center}
	\includegraphics[width=8cm,angle=270]{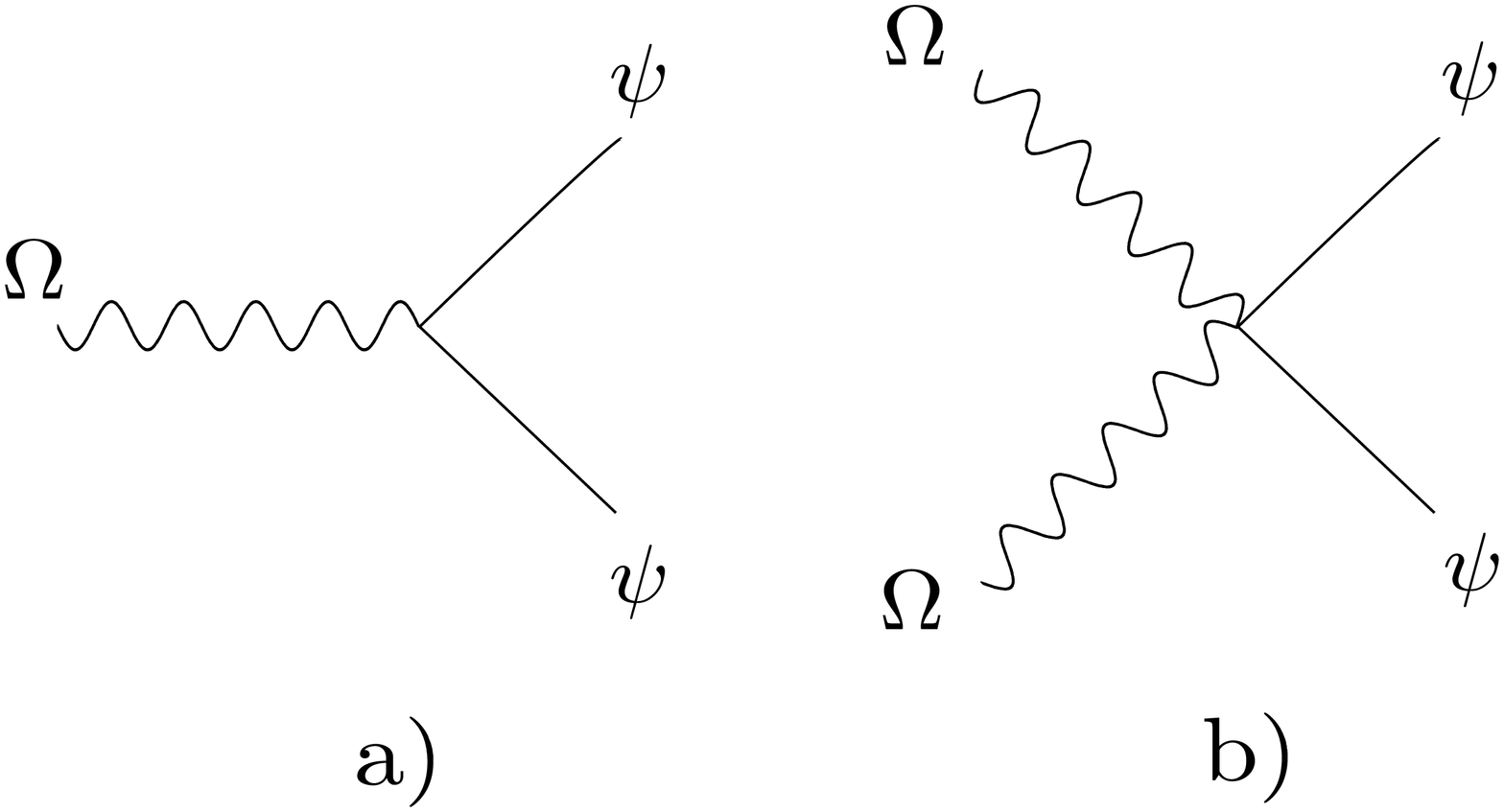}
	\end{center}
	\caption{}
	\label{fig:2}
\end{figure}

\begin{eqnarray}
\text{Fig.~\ref{fig:2} a) }&\leftrightarrow& \Omega \wedge \overline{\widetilde{\psi}}_{nc}\widetilde{\psi}_{nc} \nonumber\\
\text{Fig.~\ref{fig:2} b) }&\leftrightarrow& \frac{1}{2}\Omega^{2} \wedge \overline{\widetilde{\psi}}_{nc}\widetilde{\psi}_{nc}
\end{eqnarray}

The one loop diagrams second order in $\Omega_{\mathrm{total}}$ ``vacuum'' Feynman-diagrams are

\begin{figure}[htbp]
	\begin{center}
	\includegraphics[width=8cm,angle=270]{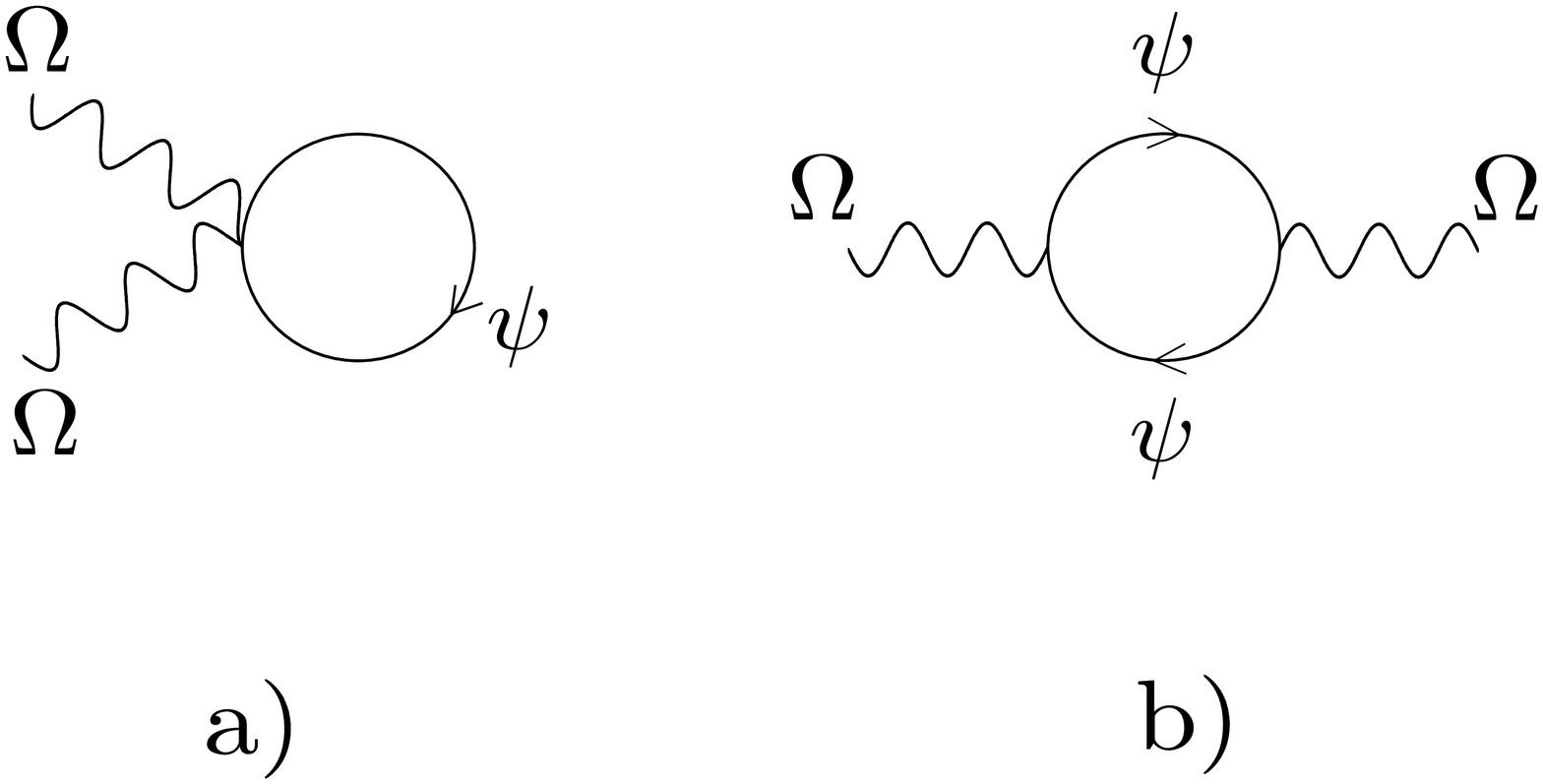}
	\end{center}
	\caption{}
	\label{fig:3}
\end{figure}

It is possible to argue that keeping included all the species proposed in our cut-off procedure in section 2 with the conditions (\ref{c1}) and (\ref{c2}) we can achieve convergence. In fact we see that for dimensional and Lorentz invariance reasons both diagrams give terms proportional to the mass square multiply a logarithmic divergence or a finite dimension as number.  Since the logarithmic divergence will not depend on the distance between $x$ and $x'$ we thus with our condition (\ref{c1}) we achieved convergence for the sum of these diagrams over the introduced species. I.e.\ the regularizations works. 
There of course should come the $\pm 1$ sign factor from the loop from the Furrey-theorem which is missing for the negatively counted species, since they are effectively bosons.
As long as we do not look for the dependence on $x-x'$ or equivalently the momentum $p$ conjugate to $x-x'$ our regularization cause cancellation of both the Fig.3 a) and Fig.3 b) diagrams. Since however Fig.3 a) is only nonzero for $x-x'=0$ it means that this diagram offer the summation over our species become totally zero.

So we only have to evaluate the diagram Fig.3 b). Now we shall remember that \\
1)this diagram only gets constructions from the \underline{extra} species (not the original fermion).\\
2)The mass(es) of the extra species are really the cut-off scale, i.e.\ very large.
Thus in the $\vec{x}$-representation this diagram \ \  only contributes when there is very short distance between the points $\vec{x}$ and $\vec{x}'$. Thus we should in principle be allowed to Taylor expand this diagram in $\vec{x}'-\vec{x}$. If we go to the Fourier transformation i.e.\ to the momentum representation we should correspondingly be allowed to take the small $\vec{p}$ approximation, only including terms proportional to the first few powers in $\vec{p}$. I.e.\ we shall assume $\vec{p}<<m$, as of course is natural since the $m$ are of the scale of the cut-off.

The diagram becomes in $\vec{x}$-representation in the simplified form of only one negatively counted species.
\\
\\
\begin{equation}
\text{Fig.~\ref{fig:3} b) }=\Omega(\vec{x})\cdot < T\bigl(G(\vec{x},\vec{x}')G(\vec{x},\vec{x}')\bigr)> \Omega(\vec{x}')
\end{equation}
where the $\vec{x}$-representation propagator for ``negatively counted particle'' is denoted 
\begin{equation}
G(\vec{x},\vec{x}')=\int \frac{d^{2}p}{(2\Pi )^{2}}\frac{i}{\not{p}-M}e^{ip(\vec{x},\vec{x}')}.
\end{equation}

Let us give a name $K(\vec{x},\vec{x}')$ to the coefficient of the product $\frac{1}{2}\Omega(\vec{x})\Omega(\vec{x}')$. Of course from translational invariance this $K(\vec{x},\vec{x}')$ only depends on the difference $\vec{x}'-\vec{x}$ and may Fourier transform its dependence on this difference $\vec{x}'-\vec{x}$.
\begin{equation}
K(\vec{x}',\vec{x})=\int e^{i\vec{p}(\vec{x}'-\vec{x})}\widetilde{K}(\vec{p})
\frac{d^{2}\vec{p}}{(2\Pi )^{2}}
\end{equation}
where then the Fourier transformed $\widetilde{K}$ of $K$ is easily seen to be given by 
\begin{equation}
\widetilde{K}(\vec{p})=M^{2} \int T_{r}\Bigl(\frac{i}{\not{q}-M}\frac{i}{\not{p}+\not{q}-M}\Bigr)
\frac{d^{2}\vec{q}}{(2\pi )^{2}}
\end{equation}

It should be had in mind that because our in the loop encircling particle is a ``negatively counted one'' we at the end do \underline{not} have the Furry-theorem's sign corresponding to the loop being a Fermion loop but rather we should treat it as a boson loop with respect to Furry sign. I.e.\ there should be no Furry-sign.

Had we included the full set of extra species the logarithmic divergence of the integral would have cancelled out to be 0. Actually for dimensional reasons $\widetilde{K}(\vec{p}=0)$ would cancel completely to zero due to (\ref{c1}). So the only important surviving term is in fact the term in $\widetilde{K}(\vec{p})$ second order in $\vec{p}$ i.e.\ we expand
\begin{equation}
\widetilde{K}(\vec{p})=\widetilde{K}(\vec{p}=\vec{0})+B\vec{p}^{2}+\ldots 
\label{6.6}
\end{equation}
and we need only to compute $B$. Here then
\begin{equation}
B=\frac{1}{2}\eta^{ij}\frac{\partial}{\partial p^{i}} \frac{\partial}{\partial p^{i}} \frac{\widetilde{K}(\vec{p})}{M^{2}}\cdot \frac{1}{d}\Bigr| _{\vec{p}=0}
\end{equation}
where $d$ is the dimension of space-time, $d=2$.

In order to perform the differentiations $\frac{\partial}{\partial p^{i}}$ with respect to the ``external'' momentum $\vec{p}$ we make use of the general rule for differentiating inverse matrices 
\begin{equation}
\frac{d}{d\xi }\bigl(\underline{\underline{A}}(\xi )^{-1}\bigr)-A(\xi )^{-1}
\frac{d\underline{\underline{A}}(\xi )}{d\xi }\underline{\underline{A}}(\xi )^{-1}
\end{equation}

(This is obtainable by differentiating first by Leibnitz rule the definitional equation for the inverse of a matrix $\underline{\underline{A}}(\xi )^{-1}\underline{\underline{A}}(\xi )=1$)
We then easily obtain
\begin{eqnarray}
B&=&M^{2}\int Tr\Bigl( \bigl( \frac{1}{\not{q}-M}\bigr)^{4}\Bigr)\frac{d^{2}q}{(2\pi )^{2}}\cdot \frac{2}{2\cdot d}\\
&=&\frac{1}{2}\int Tr\Bigl( \bigl( \frac{\not{q}+M}{q^{2}-M^{2}}\bigr)^{4}\Bigr)\frac{d^{2}q}{(2\pi )}
\end{eqnarray}

Wick-rotated this $B$ becomes, using $q^{M}_{E}=(iq^{0},q')$
\begin{equation}
B=\frac{1}{2}\int Tr\Biggl(  \frac{(\not{q}_{E}+M)^{4}}{(q^{2}_{E}+M^{2})^{4}} \Biggr) \frac{d^{2}q_{E}}{(2\pi )^{2}}
\end{equation}

We now use as usual
\begin{equation}
\not{q}^{2}_{E}=-q^{2}_{E}
\end{equation}
and
\begin{equation}
Tr(\not{q}_{E})=0.
\end{equation}
and $Tr(\underline{\underline{1}})=2$ because a Dirac spiror has 2 components in $d=2$ and also
\begin{equation}
\frac{d^{2}q_{E}}{(2\pi )^{2}}=\frac{2\pi |q_{E} |d |q_{E}|}{(2\pi)^{2}}=\frac{d(|q_{E}|^{2})}{4\pi}
\end{equation}
and obtain for the coefficient divided by $M^{2}$, $B$
\begin{equation}
B=\frac{2}{2}\int _{|q_{E}|=0}^{\infty}\frac{|q_{E}|^{4}-6|q_{E}|^{2}M^{2}+M^{4}}{(q_{E}^{2}+M^{2})^{4}}
\frac{d|q_{E}|^{2}}{4\pi}.
\end{equation}

Here we used 
\begin{equation}
(\not{q}_{E}+M)^{4}=\not{q}_{E}^{4}+4\not{q}^{3}_{E}M+6\not{q}^{2}_{E}M^{2}+4\not{q}_{E}M^{3}+M^{4}.
\end{equation}
Changing to the variable
\begin{equation}
\mu=q^{2}_{E}+M^{2}
\end{equation}
which should be integrated from $M^{2}$ to $\infty $, we write
\begin{equation}
|q_{E}|^{2}=\mu-M^{2}
\end{equation}
and thus we get 
\begin{eqnarray}
B&=&\int _{M^{2}}^{\infty}\frac{(\mu-M^{2})^{2}-6M^{2}(\mu-M^{2})+M^{4}}{\mu^{4}}\frac{d\mu}{4\pi}\nonumber\\
&=&\int _{M^{2}}^{\infty}\frac{\mu^{2}-8M^{2}\mu+8M^{4}}{\mu^{4}}\frac{d\mu}{4\pi}
\nonumber\\
&=&\int _{M^{2}}^{\infty}\Biggl[\frac{1}{\mu^{2}}-\frac{8M^{2}}{\mu^{3}}+\frac{8M^{4}}{\mu^{4}}\Biggr]
\frac{d\mu}{4\pi}
\nonumber\\
&=&\Biggl[\frac{1}{1\cdot M^{2}}-\frac{8M^{2}}{2\cdot (M^{2})^{2}}+\frac{8M^{4}}{3\cdot (M^{2})^{3}}\Biggr] \cdot \frac{1}{4\pi}
\nonumber\\
&=&\frac{1}{M^{2}}\Bigl(1-4+\frac{8}{3}\Bigr)\cdot\frac{1}{4\pi}
\nonumber\\
&=&\frac{-1}{4\pi M^{2}}\cdot\frac{1}{3}
\nonumber\\
&=&\frac{-1}{12\pi M^{2}}
\end{eqnarray}

So the coefficient in the expansion (\ref{6.6}) to $\vec{p}^{2}$ thus is 
\begin{equation}
M^{2}B=\frac{-1}{12 \pi}.
\end{equation}

Fourier transforming back to $\vec{x}'-\vec{x}$ representation we have
\begin{equation}
p^{2}- \partial_{\mu}\partial^{\mu}
\end{equation}
and we shall remember that the Ricci curvature scalar is 
\begin{equation}
R=\partial_{\mu}\partial^{\mu}\Omega _{\mathrm{TOTAL}}.
\end{equation}

Thus a term in $ln <\mathrm{sea}|e^{-i\int _{-\infty}^{\infty}Hdt}|\mathrm{sea}>$ of the form \\$ln <\mathrm{sea}|e^{-i\int _{-\infty}^{\infty}Hdt}|\mathrm{sea}>=..+\frac{1}{2}\Omega M^{2}Bp^{2}\Omega+..$, means that differentiating with respect to $\Omega$ so as to extract $T^{\mu}_{\> \> \mu}$ would give 
\begin{eqnarray}
<T^{\mu}_{\> \> \mu}>&=&M^{2}Bp^{2}\Omega \nonumber\\
&=&\frac{1}{2}M^{2}BR
\end{eqnarray}
Thus we derived
\begin{eqnarray}
T^{\mu}_{\> \> \mu}&=&\frac{1}{2}M^{2}BR=M^{2}BR\nonumber\\
&=&-\frac{1}{24\pi}R
\end{eqnarray}
This is the well-known Weyl anomaly.

\section{CONCLUSION}

We have recomputed the Weyl anomaly --- the relation of which to the conformal anomaly is described below--- in the physical picture of being due to the Dirac sea, the energy and momentum described by the $T_{\mu \nu}$- tensor is simply the ones of the Dirac sea particles. As always you can only obtain an anomaly after having had to regularize. We have proposed a somewhat new regularization method a bit reminiscent of Pauli-Villars regularization. It consists in introducing in addition to the original particles in the theory a little series of similar formal particles. The crucial feature of some of the introduced formal species of particles is that they are declared to \underline{count negatively} in the Dirac sea. Taken it that these formally introduced and negatively counted behave analogous to the original particles except that they get assigned masses of the order of the cut off scale, $M \sim \wedge$, we easily see that the contributions to e.g.\ energy density or momentum density from the numerically highest energy particles in the (combined) Dirac sea get cancelled and thus an effective cut off of these numerically high energy contributions. By a little series of formally introduced particles counted some negatively and some positively it is possible to cancel the divergencies in the Dirac sea contribution to e.g.\ the $T_{\mu \nu}$ -tensor as we discussed it in the present article. Thus one achieves in this way our regularization.

\subsection{Relation to Bosons}

Taking into account our earlier article ``Dirac sea for Bosons'' in which we consider it that there is in second quantizing Bosons removed one boson from each negative energy state analogous to Diracs adding one for fermions one sees a great similarity of our negatively counted formal regularization-particles and Bosons with the wrong spin. In fact our formally introduced negatively counted particles corresponding to the Fermions described $\psi$ above must indeed be essentially Bosons just with the wrong ``spin'' or equivalently set of field components inherited from the Fermions they shall regularize, because indeed in Feynmanloop integrals it would be needed to have from the negatively counted particles analogous loops with just the same large loop four momentum dependence of the integrand so as to cancel the divergence. Such a cancellation could just be achieved due to the Furry sign ---making the Boson and Fermion loops provided with opposite signs--- if we simply take the negatively counted specied to be opposite statistics ``for regularization purpose introduced'' particles; that is to say as bosons if the original particle species is a Fermion, or oppositely.

In this sense our regularization scheme (proposal) has some similarity to the role of super-symmetry in removing the (in)famous quadratic divergence in the Higgs        mass square thereby presenting a solution to the so called hierarchy problem. Both cancellations are due to Bosons cancelling Fermions or opposite. But there is one important difference between the SUSY cancellation in which Fermions and sfermions both obeying spin statistics theorem and our regularization method in which say a Fermion contribution is cancelled by a Boson that is a ``ghost'' in the sense that it \underline{does not obey the spin statistics theorem}.

According to our ``Dirac sea and Bosons'' work~\cite{6}~\cite{7}~\cite{8} such a violation of the spin statistics theorem would mean that our regularized theory has the possibility of negative norm states (in the Fock space) if you truly treat the to be negatively counted particles regularizing the Fermions as Bosons. But that is just as usual: a regularized theory is not satisfactory in all respects.

The Weyl symmetry anomaly we recomputed turned out that the energy momentum tensor $T^{\mu}_{\> \> \mu}$ which would have been expected to be zero in a truly Weyl invariant theory turned out to be instead
\begin{equation}
T_{\> \> \mu}^{\mu}=\frac{1}{48\pi}R.
\nonumber
\end{equation}

The easy way to see that naively the $T^{\mu }_{\> \> \mu}$ should be zero if there had been Weyl invariance meaning that varying $\Omega$ in $g_{\mu \nu}=e^{2 \Omega}\eta_{\mu \nu}$ is think of that the equation
\begin{eqnarray*}
T_{\mu \nu}&=&\frac{\partial W}{\partial g^{\mu \nu}}\nonumber\\
W&=&ln <\mathrm{sea}|e^{-i\int _{-\infty}^{\infty}Hdt}|\mathrm{sea}>
\nonumber
\end{eqnarray*}

implies $T_{\> \> \mu}^{\mu}=0$ if $W$ does not depend on $\Omega$.

Let also remark that the Weyl symmetry in which we have recomputed the anomaly is closely related to conformal invariance.

In fact if a theory in 2 dimensions is Weyl invariant, it will be conformal invariant even when a background gravity field is included. In a parametrization in which the metric tensor of the background gravity field of the form $g_{\mu \nu}=e^{2\Omega} \eta_{\mu \nu}$ a conformal transportation of the matter fields could namely be extended to the background gravity field by varying just the scalar field $\Omega(\vec{x})$. Thus if this variation, which just is a Weyl transformation $\Omega \rightarrow \Omega + \omega$ is indeed a symmetry of the theory, then the theory will also be conformally invariant even though the space-time is not flat. Then of course if there is an anomaly in the Weyl invariance symmetry there will be a corresponding one in the conformal symmetry for the \underline{curved} space.

This should also be true in general even for flat space since even in flat space a conformal transformation being extended to the gravitational field would induce a Weyl transformation. However, the special form of the Weyl anomaly, being proportional to the curvature scalar $R$ will of course turn out to vanish in the flat space. So because of this feature that the anomaly of be proportional to the curvature (scalar) $R$ one becomes allowed to say that there is in \underline{flat space} no conformal anomaly.

\subsection{Outlook}

We hope that we can find other examples of applying to anomaly calculations: 

a) the regularization method with the not spin statistics theorem obeying ``ghost'' particles

b) the Dirac sea as a physical picture.

Here we have in mind attempting to recompute the gravitational anomalies~\cite{9} in chiral Fermion theories (with Majorana-Weyl in $2+4 n$ dimensions.)

\section*{Acknowledgement}
The authors acknowledge Yasuhiro Sekino for discussions and comments. One of us (H.~B.~N.) thanks for hospitality to stay at the Niels Bohr Institute as emeritus professor. One of the authors (M.~N.) would like to thank the Niels Bohr Institute for hospitality extended to him during his stay there. M.~N. also expresses acknowledgement for financial support to the JSPS Grant-in-Aids for Scientific Research Nos.21540290, 23540332 and 24540293.



\begin{thebibliography}{99}
\bibitem{9}
  By the energy momentum tensor $T_{\mu \nu}$ we mean as usual the energy momentum tensor of the \underline{matter}, not including say gravitational field energy or momentum. However, the quantum corrections to the effective action obtain divergent terms, that can even involve second order gravity Lagrangian contributions, which thus have to be remormalized. This means that the separation between gravitational and matter energy momentum tensor becomes ambiguous. We could claim that equation (\ref{4.1}) is understood to be interpreted to be true only after the gravitational effective Lagrangian terms induced from the matter fields have been renormalized away. Strictly speaking this interpretation of (\ref{4.1}) as being valid for the ``renormalized $\langle T_{\mu \nu} \rangle_{\rm{ren}}$''  means that we take (\ref{4.1}) as a definition of $T_{\mu \nu}$ (for the matter), a definition that is needed because of the just mentioned ambiguity in the concept of ``matter''. Really are $R$ or $R^{2}$, OR tensors induced by the loop contributions from matter or gravity terms?
\end{thebibliography}
\end{document}